\documentstyle[12pt]{article}
\input amssym.def
\input amssym
\newcommand{\be}{\begin{equation}}
\newcommand{\ee}{\end{equation}}
\begin{document}

\renewcommand{\theequation}{\arabic{section}.\arabic{equation}}

\title{Modelling general relativistic perfect fluids in
field theoretic language}
\author{Nikolai V. Mitskievich\thanks{E-mail:
nmitskie@udgserv.cencar.udg.mx}\\
Physics Department, CUCEI, University of Guadalajara,\\
Guadalajara, Jalisco, Mexico}
\date{~}
\maketitle
\begin{abstract}
Skew-symmetric massless fields, their potentials being $r$-forms,
are close analogues of Maxwell's field (though the non-linear
cases also should be considered). We observe that only two
of them ($r=2$ and $3$) automatically yield stress-energy
tensors characteristic to normal perfect fluids. It is shown that
they naturally describe both non-rotating ($r=2$) and rotating
(then a combination of $r=2$ and $r=3$ fields is indispensable) 
general relativistic perfect fluids possessing every type
of equations of state. Meanwhile, a free $r=3$ field is completely
equivalent to appearance of the cosmological term in Einstein's
equations. Sound waves represent perturbations propagating on
the background of the $r=2$ field. Some exotic properties of
these two fields are outlined.
\end{abstract}

\newpage
\section{Introduction}

	Many attempts were dedicated to give a translation 
of the (semi-)pheno\-me\-no\-logical hydrodynamics to the field 
theoretical language (I use the word `translation' to contrast
with the idea of constructing a theory which could automatically
give the well known perfect fluid properties for solutions
whose physical meaning is obvious {\em ab initio}, as well as
lead to natural generalizations of old concepts). A thorough
review of many publications on Lagrangian description of
general relativistic perfect fluids is given in Brown (1993);
practically at the same time a nice paper by Carter (1994)
appeared too, which may be considered as a climax of the era
begun by Taub (1954) (probably, already even by Clebsch
(1859) in the Newtonian physics) and later developed by
Schutz (1970). One may mention few pages in (Hawking and
Ellis, 1973) on Lagrangian deduction of the dynamics of
perfect fluids, but this subject clearly served there as a
secondary accompanying theme only. Below I'll try to avoid
the translation-style approach (usually based on introduction
of several independent scalar potentials) and shall consider
another way which could be more direct and natural one. The
translation-like procedure will be used only to illustrate our
new approach in concrete examples. Except for a mere
mention of the problem of finding new exact solutions of
Einstein's equations on the basis of the proposed field
theoretical description of perfect fluids (in the concluding
Section 9), we do not touch it upon in this paper.

	The main idea is to see what simplest fields do automatically
possess the form of stress-energy tensor which is characteristic
for a perfect fluid,
\be
T^{\rm pf}=(\mu+p)u\otimes u-pg,
\ee
[see a short discussion in~(Kramer {\em et al.}, 1980), taking
into account that we use the metric $g$ with signature $+$ $-$ $-$
$-$], where $p$ is invariant pressure of the fluid, $\mu$ its
invariant mass (energy) density, and $u$ its local four-velocity.
We say `invariant' in the sense that these characteristics are
related to the local rest frame of the fluid. The `simplest' fields
are understood as those which are similar in their description
to the Maxwell one: they are massless and are described by
skew-symmetric potential tensors (of rank $r$) whose
exterior differential represents the corresponding field tensor.
Thus the connection coefficients do not enter this description.
The Lagrangian densities are functions of quadratic invariants
of the field tensors; however, some mixed invariants of the field
tensors (and sometimes, potentials) will be used, which should yield 
the same structure of the stress-energy tensor we need for a perfect
fluid (Section 2). When one speaks on a perfect fluid, its isotropy
(Pascal's property) and absence of viscosity are necessarily meant.
The most characteristic feature of the tensor (1.1) is that it has one
(single, $\mu$) and one (triple, $-p$) eigenvalues in general; this
corresponds to Pascal's property. Let us not look here after the
energy conditions; at least, a part of this problem can be ``settled''
by an appropriate redefinition of the cosmological constant to be then
extracted from the stress-energy tensor. Neither shall we consider
here thermodynamical properties of fluids, --- their phenomenological
equations of state will be used instead [see (Kramer {\em et al.},
1980)], namely the linear equation
\be
p=(\gamma-1)\mu
\ee
and the polytrope one,
\be
p=A\mu^\gamma.
\ee
Applications of these equations of state to non-rotating fluids can be
found in Sections 4 and 8 (special relativistic limit) in paragraphs
related to equations (4.4), (8.9) and (8.11).

   We shall conclude that only ranks $r=2$ and 3 correspond to (1.1),
though only the $r=2$ case leads to the $\mu+p\neq 0$ term in (1.1),
but the fluid is then non-rotating due to the $r=2$ field equations
(Sections 4 and 6); moreover, in this case one comes to a limited
class of equations of state. In the pure $r=3$ case (Section 5),
the $u\otimes u$ term in (1.1) is absent ($p=-\mu$), thus reducing
the stress-energy tensor to a pure cosmological term, the
corresponding field equation naturally yielding $\mu=\mbox{const}$.
The $r=3$ field, however, proves to be necessary alongside with
the $r=2$ one for description of rotating fluids (Section 7), as well
as of fluids satisfying more complicated equations of state
({\em e.g.}, the interior Schwarzschild solution, the end of Section
6). The scalar field case ($r=0$) does not meet some indispensable
requirements and thus should be dropped (Section 3). We give
concluding remarks in Section 9.

\section{Stress-energy tensor}

It is well known that when the action integral of a physical
system is invariant under general transformations of the
space-time coordinates, the (second) Noether theorem yields 
definitions and conservation laws of a set of dynamical
characteristics of the system. These are, in particular,
its (symmetric) stress-energy tensor and (canonical)
energy-momentum pseudotensor. The latter is important in
establishment of the commutation relations for the creation
and annihilation operators (the second quantization procedure),
while the former one acts as the source term in Einstein's
field equations. The both objects are mutually connected by 
the well known Belinfante--Rosenfeld relation. This paper is 
focused on a study of the stress-energy tensor of the ranks 2 
and 3 fields described by skew-symmetric tensor potentials (2- 
and 3-forms) whose exterior differentials serve as the corresponding 
field strengths. As it was already mentioned, this approach
does not involve the Christoffel symbols when these fields and their
interaction with gravitation are described in a coordinated basis,
thus representing the simplest scheme which resembles the
general relativistic theory of electromagnetic field.

	It is worth recalling some general definitions and relations
leading to the stress-energy tensor. Under an infinitesimal
coordinate transformation, $x'^\mu=x^\mu+\epsilon\xi^\mu(x)$,
components of a tensor or tensor density change as
$$
\delta A_a:=A'_a(x')-A_a(x)=:\epsilon 
A_a|^\tau_\sigma{\xi^\sigma}_{,\tau}
$$
(up to the first order terms; this law is, naturally, the definition
of $A_a|^\tau_\sigma$\,), $_a$~being a collective index
(the notations of Trautman~(1956), sometimes used in
formulation of the Noether theorem and general description
of covariant derivative of arbitrary tensors and tensor densities
in Riemannian geometry: $A_{a;\alpha}=A_{a,\alpha}+
A_a|^\tau_\sigma\Gamma^\sigma_{\alpha\tau}$). Then the Lie
derivative of $A_a$ with respect to a vector field $\xi$ takes form
\setcounter{equation}{0}
\be
\pounds_\xi A_a=A_{a,\sigma}\xi^\sigma-
A_a|^\tau_\sigma{\xi^\sigma}_{,\tau}\equiv A_{a;\sigma}\xi^\sigma-
A_a|^\tau_\sigma{\xi^\sigma}_{;\tau}.
\ee
The stress-energy tensor density corresponding to a Lagrangian
density ${\frak L}$, follows from the Noether theorem 
[see~(Noether, 1918; Mitskievich, 1958; Mitskievich, 1969)] as
\be
{\frak T}^\beta_\alpha:=\frac{\delta{\frak L}}{\delta g_{\mu\nu}}
g_{\mu\nu}|^\beta_\alpha\equiv
\frac{\delta{\frak L}}{\delta g^{\mu\nu}}g^{\mu\nu}|^\beta_\alpha.
\ee
Usually a rank-two tensor, and not its density, is considered,
\be
T^\beta_\alpha=(-g)^{-1/2}{\frak T}^\beta_\alpha, ~ ~
~ T^\beta_{\alpha;\beta}=0.
\ee

	Turning now to fields with skew-symmetric potentials, one
has for a rank $r$ tensor field
\be
F_{\mu\alpha\ldots\beta}:=(r+1)A_{[\alpha\ldots\beta,\mu]}
\equiv(r+1)A_{[\alpha\ldots\beta;\mu]},
\ee
where the field potential $A$ and the field tensor $F=dA$
are covariant skew-symmetric tensors of ranks $r$ and $r+1$
correspondingly, while
$$
A=\frac{1}{r!}A_{\alpha\ldots\beta}dx^\alpha\wedge\cdots\wedge
dx^\beta, ~ F=\frac{1}{(r+1)!}F_{\mu\alpha\ldots\beta}
dx^\mu\wedge dx^\alpha\wedge\cdots\wedge dx^\beta.
$$
The quadratic invariant of the field tensor is
\be
I=\ast(F\wedge\ast F)\equiv-\frac{1}{(r+1)!}
F_{\alpha_1\ldots\alpha_{r+1}}F_{\beta_1\ldots\beta_{r+1}}
g^{\alpha_1\beta_1}\cdots g^{\alpha_{r+1}\beta_{r+1}}.
\ee
[An obvious special case is the electromagnetic (Maxwell) field
($r=1$). From the expression (2.9) on, we shall use the
notations $A$ and $F$ for the potential and field tensor forms
of the electromagnetic, or $r=1$, field only, as well as $I$ for
the corresponding invariant.]

Lagrangian densities of the fields under consideration
will be taken in the general form ${\frak L}=\sqrt{-g}L(I)$,
$L(I)$ being a scalar algebraic function of the invariant
(2.5). Then relations (2.2) and (2.3) yield
\be
T^\beta_\alpha=-L\delta^\beta_\alpha-2
\frac{\partial L}{\partial g_{\mu\beta}}g_{\mu\alpha}\equiv
-L\delta^\beta_\alpha+2
\frac{\partial L}{\partial g^{\mu\alpha}}g^{\mu\beta},
\ee
so that, since $L$ depends on the metric tensor only via $I$
and due to (2.5),
\be
T^\beta_\alpha=-L\delta^\beta_\alpha-\frac{2}{s!}
\frac{dL}{dI}F_{\alpha\mu_1...\mu_s}F^{\beta\mu_1...\mu_s}.
\ee

It is easy to see that field equations can be similarly
rewritten using the function $L(I)$:
\be
\frac{\delta\frak L}{\delta A_{\alpha...\beta}}:=
\frac{\partial\frak L}{\partial A_{\alpha...\beta}}-
\left(\frac{\partial\frak L}{\partial A_{\alpha...\beta,\mu}}
\right)_{,\mu}=0 ~ \Rightarrow ~ 
\left(\sqrt{-g}\frac{dL}{dI}
F^{\alpha...\beta\mu}\right)_{,\mu}=0.
\ee
							     
	Further a more general Lagrangian density is worth being
considered,
\be
{\frak L}=\sqrt{-g}L(H,I,J,K),
\ee
a function of invariants of (skew-symmetric) fields of
ranks 0, 1, 2 and 3: 
\be
\left.\begin{array}{l}
H=\ast(d\varphi\wedge\ast d\varphi)=-\varphi_{,\alpha}
\varphi^{,\alpha}; \\
I=\ast(dA\wedge\ast dA)=-(1/2)F_{\mu\nu}F^{\mu\nu}, ~ F=dA;\\
J=\ast(dB\wedge\ast dB)=-(1/3!)G_{\lambda\mu\nu}
G^{\lambda\mu\nu}=\tilde{G}_\kappa\tilde{G}^\kappa, \\
\mbox{with } G=dB,
~ {B\!\stackrel{\mu\nu}{*}}_{;\nu}=-\tilde{G}^\mu; \\
K=\ast(dC\wedge\ast dC)=-(1/4!)W_{\kappa\lambda\mu\nu}
W^{\kappa\lambda\mu\nu}=\tilde{W}^2, ~ W=dC, 
\end{array}\right\}
\ee
where $*$ before an object is the Hodge star, 
and the duality relations hold:
\be
B\!\stackrel{\mu\nu}{*}=\frac{1}{2}E^{\mu\nu\alpha\beta}
B_{\alpha\beta}, ~ ~
G_{\lambda\mu\nu}=\tilde{G}^\kappa E_{\kappa\lambda\mu\nu}, ~
~ W_{\kappa\lambda\mu\nu}=\tilde{W}E_{\kappa\lambda\mu\nu},
\ee
$E_{\kappa\lambda\mu\nu}=\sqrt{-g}
\epsilon_{\kappa\lambda\mu\nu}$ being the Levi-Civit\`a
skew-symmetric axial tensor, while $\epsilon_{0123}=+1$.
Here $p$-forms are defined with respect to a coordinated basis
as $f=(1/p!)f_{\nu_1\nu_2\ldots\nu_p}dx^{\nu_1}\wedge
dx^{\nu_2}\wedge\cdots\wedge dx^{\nu_p}$.

As an obvious generalization of (2.2) and hence of (2.6),
the stress-energy tensor corresponding to (2.9), then 
takes the form
\be
T^\beta_\alpha=-L\delta^\beta_\alpha-
2\frac{\partial L}{\partial H}\varphi_{,\alpha}\varphi^{,\beta}-
2\frac{\partial L}{\partial I}F_{\alpha\mu}F^{\beta\mu}+
2J\frac{\partial L}{\partial J}\left(\delta^\beta_\alpha-
u_\alpha u^\beta\right)+
2K\frac{\partial L}{\partial K}\delta^\beta_\alpha
\ee
where $u_\alpha=\tilde{G}_\alpha/J^{1/2}$. When $u\cdot u=1$,
the real vector $u$ is time-like, and if imaginary, it corresponds
then to a space-like real vector. We do not consider here the
null vector case ($u\cdot u=0$).

The expressions (2.6), (2.7) and (2.12) are equivalent to those
which involve variational derivatives with respect to the
metric tensor, (2.2), if the Lagrangian density is considered as
a function of the quadratic invariants $H$, $I$, $J$ and $K$.

\section{Free (in general, nonlinear) scalar field}
\setcounter{equation}{0}
	In the free scalar field case (${\frak L}=\sqrt{-g}L(H)$),
one could also consider the (normalized) gradient of the
scalar field potential $\varphi$ as another four-velocity (say,
$\stackrel{0}{u}_\alpha=\varphi_{,\alpha}/\sqrt{|H|}$), but this
vector obviously can be timelike only if the scalar field is
essentially non-stationary (as to the four-velocity $u$ due to
the 2-form field $B$, the vector $\tilde{G}$ is automatically
timelike for stationary or static fields). In fact, the $t$-dependence
should {\em dominate} in $\varphi$, and this means that for
scalar fields normal and abnormal fluids exchange their roles
(see the next Section where these concepts are also discussed).

For the sake of completeness, we mention here the field
equation
\be
\left(\sqrt{-g}\frac{dL}{dH}\varphi^\alpha\right)_{,\alpha}=0
\ee
and the stress-energy tensor
\be
T^\beta_\alpha=-L\delta^\beta_\alpha-
2\frac{\partial L}{\partial H}\varphi_{,\alpha}\varphi^{,\beta}
\ee
of a free massless scalar field. $T^\beta_\alpha$ has then
one single and one triple eigenvalues which we denote, as
this was done for perfect fluids in (1.1), as $\mu$ and $-p$
correspondingly:
\be
\mu=2H\frac{dL}{dH}-L, ~ ~ p=L.
\ee
From these expressions we see that, if some incoherent
fluid (dust) would be described by this field, the Lagrangian
$L$ should vanish, so that the invariant $H$ has to be
(at least) constant for this solution. But then the mass density
becomes constant too, this description being obviously
applicable only to completely unphysical dust distributions.

These observations clearly show that the scalar field has to be
excluded from the list of fields suitable for description of normal
perfect fluids.

\section{Free rank 2 field}
\setcounter{equation}{0}
Let us next consider a free rank 2 field ($L$ being a function
only of $J$), thus the stress-energy tensor (2.12) reduces to 
\be
T^\beta_\alpha=\left(2J\frac{dL}{dJ}-L\right)
\delta^\beta_\alpha-
2J\frac{dL}{dJ}u_\alpha u^\beta. 
\ee
Here, $u$ evidently is eigenvector of the stress-energy tensor:
$$
T^\beta_\alpha u^\alpha=-Lu^\beta,
$$
while any vector orthogonal to $u$ is also eigenvector, this
time with the (triple) eigenvalue $2J\frac{dL}{dJ}-L$. This is 
exactly the property of the stress-energy tensor of a perfect
fluid, the only additional condition being that the vector $u$
should be a real time-like one. The latter depends however
on the concrete choice of solution of the rank 3 field equations.
Thus we come to a conclusion that 
\be
\mu=-L ~ ~ \mbox{and} ~ ~
p=L-2J\frac{dL}{dJ},
\ee
$\mu$ being invariant mass density and $p$ pressure of the
fluid. One may, of course, reinterpret this tensor as a sum
of the stress-energy tensor proper and (in general) a
cosmological term.
						 
The free field equations for the field tensor $G$ reduce to
\be
\left(J^{1/2}\frac{dL}{dJ}u_\kappa\right)_{,\lambda}=
\left(J^{1/2}\frac{dL}{dJ}u_\lambda\right)_{,\kappa} ~
\Rightarrow ~ J^{1/2}\frac{dL}{dJ}u_\lambda\equiv
\frac{dL}{dJ}\tilde{G}_\lambda=
\tilde{\Phi}_{,\lambda};
\ee
$u\cdot u=1$ by the definition. Thus the free
$r=2$ field case can describe non-rotating fluids only,
since the vector field $u$ (or, equivalently, $\tilde{G}$)
determines a non-rotating congruence. In order to
identify $u$ with the fluid's four-velocity, one has to
consider solutions with $u$ real and timelike (we call
this the normal fluid case). The null case was already
excluded from consideration, and when $\tilde{G}$ is
spacelike, one may interpret the corresponding solution
as describing a tachyonic (abnormal) fluid. The latter
notion seems to be somewhat odious, but it should be
introduced if one formulates a classification of all possible
cases of perfect fluid-like stress-energy tensors (the
well-known energy conditions are closely related to this
subject). We do not consider the tachyonic fluid case below,
moreover, we shall now show that all static spherically
symmetric solutions of the rank 2 skew-symmetric field
equations automatically yield timelike vector field $\tilde{G}$;
this should be only a part of a larger family
of physically acceptable solutions.

Perfect fluids characterized by (1.2) correspond to a
homogeneous function of $J$ as the Lagrangian, 
$L=-\sigma J^{\gamma/2}$, $\sigma>0$. The important
special cases are then: the incoherent dust ($p=0$) for
$\gamma=1$, incoherent radiation ($p=\mu/3$)
for $\gamma=4/3$, and stiff matter ($p=\mu$) for $\gamma=2$.

One may similarly treat polytropes, (1.3), though in this case
the Lagrangian is determined only implicitly. We introduce
here a notation $L=-\lambda(J)$; then $\displaystyle \mu+p
=\lambda+A\lambda^\gamma=2J\frac{d\lambda}{dJ}$ and
\be
J=\exp\left[2\int
\frac{d\lambda}{\lambda+A\lambda^\gamma}\right],
\ee
$A$ and $\gamma$ being considered as constants. It is 
clear what kind of difficulty one has to confront now: even
approximately, this relation cannot be resolved with respect
to $\lambda$, though, of course, polytropic fluids are well
described in the field theoretical language after all.
A possibility to write some function explicitly is a mere
convenience and not a necessity.

One could begin formulation of this approach with
phenomenological consideration of a perfect fluid\footnote{The
idea suggested by J. Ehlers.} just postulating the form of its
stress-energy tensor (1.1) and taking a general equation of
state in form $\mu=\mu(p)$. Define
\be
\rho=\mbox{exp}\left[\int\frac{d\mu/dp}{\mu+p}dp\right].
\ee
Then the conservation ${T^{\mu\nu}}_{;\nu}=0$ implies
$\left(\rho u^\nu\right)_{;\nu}=0$. Therefore, a
skew-symmetric tensor (superpotential) $\tilde{B}^{\mu\nu}$
should exist such that $\rho u^\mu=
{{\tilde B}^{\mu\nu}}_{\phantom{\mu\nu};\nu}$.
A direct comparison of (4.5) and (4.2) shows that
$\rho=J^{1/2}$, since, denoting 
${\displaystyle{{\tilde B}^{\mu\nu}}_{;\nu}}$
as $\tilde{G}^\mu$, we see that $J=\tilde{G}\cdot\tilde{G}=
\rho^2$ [{\em cf}. the coinciding notations in (2.10)]. This
shows that it is only natural to use a rank 2 field for
description of a perfect fluid, and the invariant $J$ is
automatically suggested; however this heuristic approach is
more closely related to the case of a Lagrangian only linearly
depending on $J$.

	In the static spherically symmetric case, with a diagonal
metric in the curvature coordinates, one has to choose
$$
B=\sin\vartheta A(r)d\vartheta\wedge d\phi, ~ ~
G=\sin\vartheta A'(r)dr\wedge d\vartheta\wedge
d\phi
$$
where the function $\sin\vartheta$ appears to make
the stress-energy tensor dependent only on the radial
coordinate; the standard spherical coordinates
notations are used. In this case,
$$
J=-A'^2\sin^2\vartheta
g^{rr}g^{\vartheta\vartheta}g^{\phi\phi}>0.
$$
For a natural 1-form basis co-moving with the fluid,
$$
\theta^{(0)}=\sqrt{g_{00}}dt=u, ~ \theta^{(1)}=\sqrt{-g_{rr}}dr, ~
\theta^{(2)}=rd\vartheta, ~ \theta^{(3)}=r\sin\vartheta d\phi,
$$
the stress-energy tensor reads
$$
T=-L(J)\theta^{(0)}\otimes\theta^{(0)}+
\left(2J\frac{dL}{dJ}-L\right)
\left(\theta^{(1)}\otimes\theta^{(1)}+
\theta^{(2)}\otimes\theta^{(2)}+\theta^{(3)}\otimes\theta^{(3)}
\right)
$$
in conformity with (4.2).

\section{Free rank 3 field. The only interpretation:
cosmological term}
\setcounter{equation}{0}
In this case the Lagrangian depends only on the invariant $K$;
thus
\be
T^\beta_\alpha=\left(2K\frac{dL}{dK}-L\right)
\delta^\beta_\alpha=-\frac{\Lambda}{\varkappa}\delta^\beta_\alpha,
\ee
$\varkappa$ being Einstein's gravitational constant.
This stress-energy tensor is merely proportional to the metric
tensor; therefore the coefficient $2K\frac{dL}{dK}-L=
-\Lambda/\varkappa$ obviously should be constant. It is trivially
constant (and equal to zero) indeed when $L\sim K^{1/2}$,
the field components $W^{\kappa\lambda\mu\nu}$ being then
arbitrary. Otherwise, it becomes constant (and nonzero)
due to the field equations to which vanishing of the stress-energy
tensor divergence is equivalent. Indeed, the equations
\be
\left(\sqrt{-g}\frac{dL}{dK}
W^{\kappa\lambda\mu\nu}\right)_{,\nu}=0
\ee
reduce to
\be
K^{1/2}\frac{dL}{dK}=\mbox{const.}
\ee
since $\sqrt{-g}E^{\kappa\lambda\mu\nu}=
-\epsilon_{\kappa\lambda\mu\nu}=$ const. We see that the both
cases (when $L\sim K^{1/2}$ and $L\nsim K^{1/2}$)
exactly correspond to the above conclusions. In the first case
this does not deserve comments, but when $L\nsim K^{1/2}$,
the left-hand side expression in (5.3) is really a function
of $K$. Hence from (5.3) it follows that $K$ itself should be
constant. Thus the `cosmological constant' $\Lambda$ which 
appears in (5.1), is really constant due to the field equations.
These equations, in a sharp contrast to the usual equations
of mathematical physics, cannot be characterized as hyperbolic
ones (or else). Moreover, the case of $L\sim K^{1/2}$ corresponds
to vanishing of the cosmological constant, and the field
equations do now impose no conditions on $K$ whatsoever ---
the rank 3 field is then {\em arbitrary} due to the field equation,
a very specific situation for the field theory indeed!

If $L=\sigma K^k$  with a positive constant $\sigma$, then 
$2k<1$ corresponds to the de~Sitter case; $2k=1$, to
the absence of cosmological constant (this is the case of
a {\em phantom} rank 3 field which is completely arbitrary,
and else, it does not produce any stress-energy tensor
at all); finally, $2k>1$ corresponds to the anti-de~Sitter
case [see for standard definitions (Hawking and Ellis, 1973)].
We propose to call the rank 3 field a cosmological field;
another --- Machian --- reason for this will become obvious
after a consideration of rotating fluids.

\section{Non-rotating fluids}
\setcounter{equation}{0}
In a comoving frame, the local four-velocity of a fluid is
$u^\mu\sim \delta^\mu_0$, and the $x^0$ coordinate lines
should form a non-rotating congruence.
Since $u\cdot u=1$, in the case of a normal fluid,
\be
u^\mu=\delta^\mu_0/\sqrt{g_{00}}, ~ ~ \tilde{G}^\mu=
\Xi\delta^\mu_0,
\ee
$\Xi$ being a function of the four (in general) coordinates.
Thus 
\be
J=\Xi^2g_{00}, ~ \mbox{and} ~ u^\mu=\tilde{G}^\mu/\sqrt{J}.
\ee
To be more concise, we shall consider here the case of a
homogeneous function $L(J)=\sigma J^k$.
Then $J^{k-1}\tilde{G}_\lambda=\tilde{\Phi}_{,\lambda}$, 
$\tilde{\Phi}$ being a pseudopotential (with the pseudoscalar
property). 

Let us now consider some perfect fluid solutions in general
relativity [for excellent reviews see (Kramer {\em et al.}, 
1980; Delgaty and Lake, 1998)]. It is convenient to write
this solution in comoving coordinates. Moreover, let the
fluid satisfy an equation of state $p=(2k-1)\mu$ with
$k=$ const. Apart from the metric coefficients, there will
be only one independent function characterizing the fluid
(and its motion), say, $\mu$. In the scheme outlined above,
this function should be related to $\Xi$, the only
independent function involved in the $r=2$ field (the metric
tensor is supposed to be the same in the perfect fluid and
$r=2$ field languages). Clearly, the problem then reduces
to a determination of the relationship between the two
functions. One finds immediately 
\be
\mu=\sigma J^k, ~ \mbox{thus} ~
\Xi=(\mu/\sigma)^{1/2k}/\sqrt{g_{00}}.
\ee
Hence,
\be
\tilde{G}^\mu=\frac{1}{\sqrt{g_{00}}}(\mu/\sigma)^{1/2k}
\delta^\mu_0 ~ \mbox{and} ~ \tilde{\Phi}_{,\mu}=
k(\mu/\sigma)^{(2k-1)/2k}g_{0\mu}/\sqrt{g_{00}}.
\ee
\subsubsection*{Example: the Klein metric}

The Klein metric~(Klein, 1947; Kramer {\em et al.}, 1980) 
describes a static space-time filled with incoherent radiation,
$p=\mu/3$. In this case,
$$
ds^2=rdt^2-\frac{7}{4}dr^2-r^2\left(d\vartheta^2+
\sin^2\vartheta d\phi^2\right),
$$
$$
\mu=3/(7\varkappa r^2), ~ ~ k=2/3.
$$
Then, obviously,
$$
\Xi=\left(\frac{3}{7\varkappa\sigma}\right)^{3/4}\frac{1}{r^2},
~ ~ \tilde{\Phi}=\frac{2}{3}\left(\frac{3}{7\varkappa\sigma}
\right)^{1/4}t.
$$

\subsubsection*{Example: the Tolman--Bondi solution~(Tolman,
1934; Bondi, 1947)}

Now, 
$$
ds^2=d\tau^2-\exp(\lambda(\tau,R))dR^2-r^2(\tau,R)
\left(d\vartheta^2+\sin^2\vartheta d\phi^2\right),
$$
$$
\mu=\frac{F'}{\varkappa r'r^2}, ~ ~ p=0, ~ ~ k=1/2, 
$$
$$
r=\frac{F}{2f}(\cosh\eta-1), ~ ~ 
\sinh\eta-\eta=\frac{2f^{3/2}}{F}(\tau_0-\tau),
$$
$F$, $f$ and $\tau_0$ being arbitrary functions of $R$.
The translation into the $s=2$ field language reads simply
$$
\Xi=\frac{F'}{\varkappa\sigma r'r^2}, ~ ~ \tilde{\Phi}=\tau.
$$

It is equally easy to cast the Friedmann--Robertson--Walker
cosmological solutions in the rank 2 field form (in fact,
the FRW universe filled with an incoherent dust represents a
special case of the Tolman--Bondi solution).

\subsubsection*{Example: the interior Schwarzschild
solution~(Kramer {\em et al.}, 1980)}

The interior Schwarzschild solution is now a special
case to be treated in more detail. Its characteristic feature
is that the mass density of the fluid with which
it is filled, is constant, while the fluid's pressure
decreases when the radial coordinate grows, vanishing on
some spherical boundary (thus making it
possible to join this solution with the exterior vacuum
region). However this property clearly contradicts to
the relation between $\mu$ and $p$ obtainable from a
Lagrangian depending on one invariant, $J$, only. Therefore
one has to consider interaction, say, of $r=2$ and $r=3$
fields. We choose the corresponding Lagrangian to be
$L(J,K)=-M(J)\left(1-\alpha K^{1/2}\right)$ (the rank 3 field
obviously being a phantom one). Then 
$$
T^\beta_\alpha=\left[M(J)-2J\frac{dM}{dJ}\left(1-\alpha
K^{1/2}\right)\right]\delta^\beta_\alpha+
2J\frac{dM}{dJ}\left(1-\alpha K^{1/2}\right)u_\alpha u^\beta,
$$
hence the former expression for $\mu$ is not changed,
but pressure is now a function of the invariant $K$ 
{\em arbitrarily} depending on coordinates:
\be
\mu=M(J), ~ ~ p=2J\frac{dM}{dJ}\left(1-\alpha K^{1/2}\right)
-M(J).
\ee

The fact that $K$ really may be chosen arbitrarily, follows
from the field equations. For the $r=2$ field one has
\be
d\left[\frac{dM}{dJ}\left(1-\alpha K^{1/2}\right)
\tilde{G}\right]=0,
\ee
and for the $r=3$ field,
\be
M(J)=\mbox{const.},
\ee
without any other conditions on $K$. The latter equation is 
exactly what we needed, and the first one then reduces to
$d\left[\left(1-\alpha K^{1/2}\right)\tilde{G}\right]=0$ or, in
the static case when $K$ is independent of $x^0$ and
$\tilde{G}=J^{1/2}\sqrt{g_{00}}dx^0$, simply to
\be
\left(1-\alpha K^{1/2}\right)\sqrt{g_{00}}=q^2,
\ee
$q$ being a constant. 

However, the last equation seems to impose a critically
strong restriction on the choice of $K$ (yet having been
arbitrary) which should now
automatically fit the expression for pressure. Let us
see if this is the case for the interior Schwarzschild
solution. The latter is described by
\be
\left.
 \begin{array}{l}
\displaystyle{
ds^2=\left(a-b\sqrt{1-\frac{r^2}{R^2}}\right)^2\!\!dt^2-
\frac{dr^2}{1-r^2/R^2}-r^2\!\left(d\vartheta^2+
\sin^2\vartheta d\phi^2\right),}
\\
\displaystyle{\mu=\frac{3}{\varkappa R^2}, ~ ~ 
p=\frac{3}{\varkappa R^2}\!\left(\frac{2a}{3\sqrt{g_{00}}}
-1\right),}
\end{array}\right\}
\ee
$a$, $b$ and $R$ being constants [see for
details~(Kramer {\em et al.} 1980)]. If we take $M=\sigma J^k$,
it is readily found that all conditions are satisfied indeed
for $k=a/3q$. Then for $\mu= $const. it is always possible
to consider a linear $r=2$ field, $k=1$: one has only to
choose $q=a/3$.

\section{Rotating fluids}
\setcounter{equation}{0}
We came to conclusions that the $r=2$ and $r=3$ fields have 
stress-energy tensors possessing eigenvalues typical to 
perfect fluids: in the free field cases, the $r=2$ field with 
the eigenvalues characteristic for a usual isotropic perfect fluid,
and the $r=3$ field, with only one quadruple eigenvalue (thus
the stress-energy tensor is proportional to the metric tensor:
the cosmological term form). For description of a perfect fluid
with the equation of state $p=(\gamma-1)\mu$ and a given
constant value of $\gamma$ one needs only one function, say,
the mass density $\mu$ (the metric tensor is considered as
already given, and the system of coordinates is supposed to
be co-moving with the fluid, thus the four-velocity vector is
$u^\mu=(g_{00})^{-1/2}\delta^\mu_0$). It seemed that this
situation in all cases fits well for translating into
the $r=2$ field language. But we were confronted with the no
rotation condition for perfect fluid when the rank 2 field
was considered to be free. It is clear that the only remedy is
in this case an introduction of a non-trivial source term
in the $r=2$ field equations, thus a change to the non-free
field case or, at least, to include in the Lagrangian a
dependence on the rank 2 field potential $B$.

The simplest way to do this is to introduce in the Lagrangian
density dependence on a new invariant $J_1=
-B_{[\kappa\lambda}B_{\mu\nu]}B^{[\kappa\lambda}B^{\mu\nu]}$
which does not spoil the structure of stress-energy tensor, 
simultaneously yielding a source term (thus permitting to 
destroy the no rotation property) without changing the 
divergence term in the $r=2$ field equations. We shall use
below three invariants: the obvious ones, $J$ and $K$, 
and the just introduced invariant of the $r=2$ field 
{\em potential}, $J_1$. One easily finds that
\be
B_{[\kappa\lambda}B_{\mu\nu]}=-\frac{2}{4!}B_{\alpha\beta}
B\!\stackrel{\alpha\beta}{\ast}E_{\kappa\lambda\mu\nu}
\ee
where $B\!\stackrel{\alpha\beta}{\ast}:=\frac{1}{2}B_{\mu\nu}
E^{\alpha\beta\mu\nu}$ (dual conjugation). Thus
$J_1^{1/2}=6^{-1/2}B_{\alpha\beta}
B\!\!\stackrel{\alpha\beta}{\ast}$. In fact, $J_1=0$, if $B$ is
a simple bivector ($B=a\wedge b$, $a$ and $b$ being
1-forms; only the four-dimensional case to be considered);
this corresponds to all types of rotating fluids discussed in
existing literature. This {\em cannot however annul} the
expression which this invariant contributes to the $r=2$
field equations: up to a factor, it is equal to
$\partial J_1^{1/2}/\partial B_{\mu\nu}\neq 0$. 
Thus let the Lagrangian density be
\be
{\frak L}=\sqrt{-g}(L(J)+M(K)J_1^{1/2}).
\ee

The $r=2$ field equations now take the form ({\em cf.} (4.3))
\be
d\left(\frac{dL}{dJ}\tilde{G}\right)=
\sqrt{2/3}M(K)B,
\ee
which means that introduction of rotation of the fluid
destroys the gauge freedom of the $r=2$ field.
In their turn, the $r=3$ field equations ({\em cf.} (5.2)
and (5.3)) yield the first integral
\be
J_1^{1/2}K^{1/2}\frac{dM}{dK}=\mbox{const}\equiv 0
\ee
(when $J_1=0$, as it was just stated).
It is obvious that $K$ (hence, $M$) {\em arbitrarily} depends
on the space-time coordinates, if only the $r=3$ field
equations are taken into account. Though the $r=2$ field
equations (7.3) apparently show that the $\tilde{G}$ congruence
should in general be rotating, the $r=2$ field $B$ is an exact
form for solutions with constant $M(K)$, thus its substitution
into the left-hand side of (7.3) via $\tilde{G}$ leads
trivially to vanishing of $G$ (and hence $B$). Hence in
a non-trivial situation the cosmological field $K$
(see (2.10)) has to be essentially non-constant.

But the complete set of equations contains Einstein's
equations as well. One has to consider their sources and
the structure of their solutions (some of which fortunately
are available) in order to better understand this remarkable
situation probably never encountered in theoretical physics
before.

The stress-energy tensor which corresponds to the new Lagrangian
density (7.2), is
\be
T^\beta_\alpha=\left(-L-MN+2J\frac{dL}{dJ}+2KN\frac{dM}{dK}+
2J_1M\frac{dN}{dJ_1}\right)\delta^\beta_\alpha-
2J\frac{dL}{dJ}u_\alpha u^\beta
\ee
where we have used $N(J_1)=J^{1/2}_1$. It is obvious that
only the terms involving $L$ and $J$ survive here
($J_1=0=N$). For a perfect fluid with the equation of state
$p=(\gamma-1)\mu$, one finds $L=-\sigma J^{\gamma/2}$, thus 
$T^\beta_\alpha=-\gamma Lu_\alpha u^\beta+(\gamma-1)L
\delta^\beta_\alpha$.

Then one has a translation algorithm between the
traditional perfect fluid and $r=2$ field languages:
\be
\left. \begin{array}{l}
\displaystyle
{\mu=-L=\sigma J^{\gamma/2}, ~ ~ \tilde{G}^\mu
=\Xi\delta^\mu_t,
~ ~ \Xi=\frac{1}{\sqrt{g_{00}}}\left(\frac{\mu}{\sigma}
\right)^{1/\gamma}},\\
\displaystyle{G=dB=\sqrt{3/2}
d\left(\frac{1}{M(K)}\right)\wedge d\left(
\frac{dL}{dJ}\tilde{G}\right)}
\end{array}  \right\}
\ee
({\em cf.} (7.3)). The function $M$ depends arbitrarily on
coordinates; thus one can choose its adequate form using
the last relation without coming into contradiction with
the dynamical equations.

We see that the cosmological field $K$ plays a very special
role in description of rotating fluids. This field makes it
possible to consider rotation, but its own field equations
do not impose any restriction on $K$. (A similar situation,
but without rotation, was observed above in the case of the
interior Schwarzschild solution.) In each case, one has to
adjust the $K$ field using the gravitational field solutions,
thus from global considerations (this being the final analysis
of considerations of the last paragraphs). Together with the
fact that the free $K$ field results in introduction of the
cosmological constant, these properties of the cosmological
field recall the ideas of the Mach principle and a practically
forgotten hypothesis due to Sakurai (1960).

\subsubsection*{Example: The G\"odel universe~(G\"odel, 1949)}

G\"odel's universe filled with rotating perfect fluid is
described by
$$
ds^2=a^2\left(dt^2+2\sqrt{2}zdtdx+z^2dx-dy^2-z^{-2}dz^2\right),
~ \sqrt{-g}=a^4,
$$
$$
p=\mu=\frac{1}{2\varkappa a^2}, ~ ~ u^\mu=a^{-1}\delta^\mu_t, ~
~ k=1 ~ ~ (\gamma=2)
$$
[in the book~(Kramer {\em et al.}, 1980) $p$ and $\mu$ take other
values, since a cosmological term is there considered,
but this is only a matter of convention; moreover,
there are misprints in the book: the factor $a^2$ should
be put in the denominator, as we have written above].
Now it is easy to find
$$
\Xi=\left(2\varkappa\sigma a^4\right)^{-1/2}, ~ ~
\tilde{G}=(2\varkappa\sigma)^{-1/2}\left(dt+
\sqrt{2}z\,dx\right), ~ ~ J=\left(2\varkappa\sigma a^2
\right)^{-1},
$$
while $G=a^2(2\varkappa\sigma)^{-1/2}dx\wedge dy\wedge dz$.
Hence (see also (7.3))
$$
d\left(\frac{dL}{dJ}\tilde{G}\right)=
\sqrt{\sigma/\varkappa}dx\wedge dz=\sqrt{2/3}M(K)B,
$$
so that
$$
G=dB=\sqrt{\frac{3\sigma}{2\varkappa}}d\left(\frac{1}{M}\right)
\wedge dx\wedge dz.
$$
This gives $\displaystyle{M=-\frac{\sqrt{3}\sigma}{a^2y}}$
and 
$\displaystyle{B=\frac{a^2y}{\sqrt{2\varkappa\sigma}}
dx\wedge dz}$.

\subsubsection*{Example: Davidson's fluid~(Davidson, 1996)}

Another stationary solution with fluid being in a certain
sense in a rigid body rotation, is described by the metric
$$
ds^2=P\left(dt+\sqrt{23/8}\,ar^2d\phi\right)^2-r^2P^3d\phi^2
-P^{-3/4}\left(dr^2+dz^2\right),
$$
$\sqrt{-g}=rP^{5/4}$, while
$$
P=\sqrt{1+a^2r^2}, ~ ~ \gamma=5/3, ~ ~ 
\mu=\frac{9a^2}{2\varkappa}P^{-5/4}.
$$
We find
$$
\Xi=\left(\frac{9a^2}{2\varkappa\sigma}\right)^{3/5}P^{-5/4},
~ ~ J=\left(\frac{9a^2}{2\varkappa\sigma}\right)^{6/5}P^{-3/2},
$$
$$
\frac{dL}{dJ}\tilde{G}=-\frac{5\sigma}{6}\left(
\frac{9a^2}{2\varkappa\sigma}\right)^{2/5}\!\!\left(dt+
\sqrt{23/8}\,ar^2d\phi\right), ~ ~ 
G=\left(\frac{9a^2}{2\varkappa\sigma}\right)^{3/5}\!\!r\,dr\wedge
dz\wedge d\phi,
$$
$$
M=\left(
\frac{9a^2}{2\varkappa\sigma}\right)^{-1/5}
\frac{5\sqrt{23}\sigma a}{4\sqrt{3}z}, ~ ~
B=-\left(\frac{9a^2}{2\varkappa\sigma}\right)^{3/5}
\! zr\,dr\wedge d\phi.
$$

In both of these examples we have determined $M$ as a
function of a coordinate, without mentioning the $r=3$ field
tensor, since in the rotating perfect fluid theory the
coordinates dependence of $M$ only matters. It is clear
that our considerations are in a complete agreement with
the field equations. 

\section{Special relativistic theory}
\setcounter{equation}{0}
In special relativity, when $g_{\mu\nu}=\eta_{\mu\nu}=
\mbox{diag}(1,-1,-1,-1)$ (in Cartesian coordinates),
one does not use Einstein's equations, so that a
homogeneous distribution of a perfect fluid in infinite
flat space-time becomes admissible. We shall consider
here the behaviour of weak perturbations on the background
of such a homogeneous field of a non-rotating perfect fluid.
Then in the zeroth approximation $\tilde{G}$ coincides with
the four-velocity of the fluid, $u=dt$ (in co-moving
coordinates; $t=x^0$), $J=1$ (the background situation).

Now let a perturbation be introduced, thus
\be
\tilde{G}^\kappa=\delta^\kappa_t+\delta\tilde{G}^\kappa, ~
J=1+2\delta\tilde{G}^t+\delta\tilde{G}^\kappa
\delta\tilde{G}_\kappa.
\ee
These relations might be considered as exact ones, though
it is easy to see that, if one does not intend to consider
the linear approximation only, it would be worth 
expressing the very $\delta\tilde{G}$ as a series of
terms which describe all orders of magnitude of the
perturbations. However in the present context this will
be of minor importance, and we shall deal with linear
terms only. Then
\be
L(J)=L(1)+2\left[\frac{dL}{dJ}\right]_1\delta\tilde{G}^t+
...;
\ee
here the points denote higher-order terms. The expression
of $L(J)$ is equivalent (up to its sign) to the mass
density, but one has still to take into account the field
equations (4.3). These read, in similar notations,
\be
\tilde{\Phi}_{,\kappa}=\left[\frac{dL}{dJ}\right]_1
\delta^t_\kappa+\left[\frac{dL}{dJ}\delta^\lambda_\kappa+
2\frac{d^2L}{dJ^2}\delta^t_\kappa\delta^\lambda_t\right]_1
\delta\tilde{G}_\lambda+...\; .
\ee
The only property which matters in this expression, is
its gradient form. We arrive to the following two equations
(the Latin indices being three-dimensional),
\be
\left(\tilde{\Phi}\right)_{,t,i}=
\left(\tilde{\Phi}\right)_{,i,t} ~ \Rightarrow ~
\left[\frac{dL}{dJ}+
2\frac{d^2L}{dJ^2}\right]_1\left(\delta\tilde{G}_t\right)_{,i}=
\left[\frac{dL}{dJ}\right]_1\left(\delta\tilde{G}_i\right)_{,t}
\ee
and
\be
\left(\tilde{\Phi}\right)_{,i,j}=
\left(\tilde{\Phi}\right)_{,j,i} ~ \Rightarrow ~
\left[\frac{dL}{dJ}\right]_1
\left(\delta\tilde{G}_i\right)_{,j}=
\left[\frac{dL}{dJ}\right]_1\left(\delta\tilde{G}_j
\right)_{,i}.
\ee
One has to conclude that this set of equations is
satisfied if
\be
\delta\tilde{G}_i=\left[\frac{dL/dJ+2d^2L/dJ^2}{dL/dJ}
\right]_1\left(\int\delta\tilde{G}_tdt+\phi(\vec{x})
\right)_{,i},
\ee
with two still non-determined functions, 
$\delta\tilde{G}_t(t,\vec{x})$ and $\phi(\vec{x})$.
But we did not yet taken into account that
$\delta\tilde{G}$ (as well as $\tilde{G}$) is
divergenceless. This actually means that
$$
\delta\tilde{G}^t_{,t}=-\delta\tilde{G}^i_{,i}=
\delta\tilde{G}_{i,i}=\left[\frac{dL/dJ+2d^2L/dJ^2}{dL/dJ}
\right]_1\Delta\left(\int\delta\tilde{G}_tdt+\phi(\vec{x})
\right) ,
$$
$\Delta$ being the Laplacian operator. Differentiating
the both sides of this relation with respect
to $t=x^0$, we find at last
\be
\frac{\partial^2\delta\tilde{G}_t}{\partial t^2}=
\left[\frac{dL/dJ+2d^2L/dJ^2}{dL/dJ}
\right]_1\Delta\delta\tilde{G}_t,
\ee
a modification of the D'Alembert equation (involving a
velocity different from that of light). Since
propagation properties of perturbations
of the mass density $\mu$, of the Lagrangian $L$ and of
the field component $\tilde{G}_t$ mutually coincide in
the first approximation, one has to conclude that the
velocity of the low amplitude
density (sound) waves in a fluid is equal to
\be
c_{\rm s}=\sqrt{\left[\frac{dL/dJ+2d^2L/dJ^2}{dL/dJ}\right]_1}
\ee
in units of the velocity of light. One has, of course, to
remember that in this theory the laws of thermodynamics
were used only implicitly (via equations of state). However
some important properties of the sound waves already can be
seen in this result.

Let us consider first the simplest case which is described
by the equation of state (1.2). Then $L=-\sigma J^{\gamma/2}$,
and we have
\be
c_{\rm s}=\sqrt{\gamma-1}.
\ee
When $\gamma=1$, the perturbations do not propagate
(in the co-moving frame of the fluid); this is the
case of an incoherent dust whose particles interact
only gravitationally, {\em i.e.} do not interact in
a theory devoid of gravitation (special relativity).
When $\gamma=2$, we have a stiff matter, in which
(as it is well known) sound propagates with the
velocity of light, and this is exactly the case in
our field theoretical description: $c_{\rm s}=1$.
When the value of $\gamma$ lies between 1 and 2,
we have more or less realistic fluids, the
velocity of sound in them being less than that of
light. For example, in the case of incoherent
radiation (see a consideration of the Klein metric
above), $c_{\rm s}=1/3$.

Turning to consideration of a polytrope (1.3) and
taking into account its field theoretical description
(4.4), it is easy to find for the sound velocity (8.8)
the corresponding form
\be
c_{\rm s}=\sqrt{\left[1-2\left(\frac{dJ}{dL}\right)^{-2}
\frac{d^2J}{dL^2}\right]_1}
\ee
or, after a substitution of (4.4), exactly the standard
expression
\be
c_{\rm s}=\sqrt{\gamma p/\mu}.
\ee

It is worth stressing that in this section all considerations
were only restricted to absence of gravitational field as
well as to weak perturbations of the fluid density, but the
velocity of propagation of the perturbations may be
relativistic one. Thus the standard expression (8.11)
represents in fact an exact generalization of $c_{\rm s}$
to the relativistic case; similarly, (8.9) gives correct
value of the velocity of sound in ultrarelativistic cases
important in astrophysical context.

\section{Concluding remarks}

As a summary of the just described results and in anticipation
of some others (to be presented elsewhere), it is worth
systematizing the present approach in the 3+1-dimensional
spacetime. Our conclusions are essentially based on a
consideration of the stress-energy tensor of $r$-form fields
($r=0,$ 1, 2 and 3), the fact which makes it clear why these
conclusions merely partially coincide with those of Weinberg
(1996, Section 8.8) where only the gauge covariance
properties are taken into account.

A field whose potential is a skew-symmetric tensor of rank 4
(being identically a closed form in four dimensions), has
only trivial field strength tensor thus leaving for
consideration the four fields used in (2.12).

The rank 3 field does not correspond to any real quantum
particles (a result obtained in collaboration with
H. Vargas Rodr{\'\i}guez, to be published elsewhere), thus
these particles should be only virtual ones. In the classical
theory, the rank 3 field with any degree of non-linearity is
equivalent to appearance of cosmological constant in
Einstein's equations; when the Lagrangian density is
proportional to $K^{1/2}$, the cosmological constant vanishes
(thus suggesting a new interpretation of the very fact). 
The global nature of Mach's principle (admittedly related to
rotation phenomena) also seems to justify consideration of
the rank 3 field on a basis similar to that of the
hypothetical fundamental cosmological field proposed by
Sakurai (1960).

The rank 2 field describes (sometimes in interaction with
the cosmological field) perfect fluids. The second
quantization of the free rank 3 field yields real quanta,
but they have only spin zero: all other particles appear as
thoroughly virtual ones (another result in collaboration
with Vargas Rodr{\'\i}guez, also not included in this paper).

Then comes the rank 1 field which, in its linear case, is
the Maxwellian one, making all commentaries unnecessary.
And the last is the scalar field; I would add here (to the
information given in Sections 2 and 3) only one comment
on this field: its interaction with the rank 2 field mimics the
electromagnetic field, thus exactly and with the same
degree of simplicity reproducing, for example, the
Reissner--Nordstr\"om black hole spacetime without any
electromagnetic field whatsoever (Mitskievich, 1998).
This all follows from the stress-energy tensor (2.12).

It is worth mentioning that in the 2+1-dimensional spacetime
the $r=1$ field, formerly, the (non-linear) Maxwell one,
now describes perfect fluids, while the $r=2$ field is
responsible for the cosmological term in 3D Einstein's
equations.

The proposed description of perfect fluids is simple, and
it yields exactly the same characteristics of perfect
fluids and relations between these characteristics which
are already well established in the other approaches
(see, {\em e.g.}, our consideration of the special
relativistic limit of the theory, yielding the properties
of sound waves in fluids).
Moreover, our description suggests (and simplifies the
realization of) some new lines in generalization of the
theory of perfect fluid (due to an extensive use of the
Lagrangian formalism), in particular, it makes the second
quantization of (the sound in) the perfect fluid in fact
a mere routine.

The use of standard field theoretical methods for description
of perfect fluids and their excitations (phonons), may also
help in evaluation of an effect of {\v C}erenkov-type
radiation of sound by narrow-fronted gravitational wave
jets (or, gravitons) in matter. Another possible application
of the proposed description of perfect fluids may be related
to construction of exact Einstein--Euler fields (gravitation
and perfect fluid) using the properties of Killing--Yano
tensors, if these would be admitted by the vacuum seed
spacetimes ({\em cf.} the method proposed in (Horsk\'y and
Mitskievich, 1989) for Einstein--Maxwell fields which uses
Killing vectors). 

\section*{Acknowledgements}
	This work was partially supported by the CONACyT
(Mexico) Grant 1626P-E, by a research stipend of the 
Albert-Einstein-Institut (Potsdam, Germany) and by a travel
grant of the Universidad de Guadalajara.

	My sincere thanks are due to B. Carter for kind attention
and valuable information. I am greatly indebted to the colleagues
and administration of the Albert-Einstein-Institut 
(Max-Planck-Institut f\"ur Gravitationsphysik) in Potsdam,
Germany, for friendly hospitality during my stay there in
March-April of 1997. I am grateful to K. Bronnikov, R. Meinel,
V. Melnikov, G. Neugebauer, N. Sali\'e, E. Schmutzer, B. Schutz
and all colleagues for constructive discussions in seminars
of the Albert-Einstein-Institut and the Universit\"at Jena.
My special thanks are due to J. Ehlers who not only
friendly encouraged and helped me in many important cases,
but also did suggest fruitful concrete ideas some of which
I used in this paper. I gratefully acknowledge helpful remarks
of V. Efremov and H. Vargas Rodr{\'\i}guez of the Universidad
de Guadalajara.

\newpage

\section*{References}
~~~~~

	Bondi, H. (1947). {\em Mon. Not. Roy. Astr. Soc.},
{\bf 107}, 410.

	Brown, J.D. (1993). {\em Class. Quantum Grav.},
{\bf 10}, 1579.

	Carter, B. (1989). In: {\em Lecture Notes in
Mathematics}, {\bf 1358}, ed. A.~Anile and Y.~Choquet-Bruhat,
pp 1--64, Springer Verlag, Heidelberg.

	Carter, B. (1994). {\em Class. Quantum Grav.} ,
{\bf 11}, 2013.

	Clebsch, A. (1859). {\em J. reine angew. Math.},
{\bf 56}, 1.

	Davidson, W. (1996). {\em Class. Quantum Grav.},
{\bf 13}, 283.

	Delgaty, M.S.R., and Lake, K. (1998). {\em Los Alamos
Preprint gr-qc/9809013}, 25 pp.

	G\"odel, K. (1949). {\em Rev. Mod. Phys.},
{\bf 21},  447.

	Hawking, S.W., and Ellis, G.F.R. (1973).
{\em The Large Scale Structure of Space-time}, Cambridge
University Press, Cambridge.

	Horsk\'y, J., and Mitskievich, N.V. (1989). {\em Czech. J.
Phys.}, {\bf B39}, 957.

	Klein, O. (1947). {\em Ark. Mat. Astr. Fys.},
{\bf A 34}, 1.

	Kramer, D., Stephani, H., MacCallum, M., and
Herlt, E. (1980). {\em Exact Solutions of Einstein's Field Equations},
DVW, Berlin/Cambridge University Press, Cambridge.

	Misner, Ch.W., Thorne, K.S., and Wheeler, J.A.
(1973). {\em Gravitation}, W.H.~Freeman, Los Angeles.

	Mitskievich (Mizkjewitsch), N.V. (1958).
 {\em Ann. Phys. (Leipzig)}, {\bf 1}, 319.

	Mitskievich, N.V. (1969). {\em Physical 
Fields in General Relativity}, Nauka, Moscow (in Russian).

	Mitskievich, N.V. (1998). {\em Gravitation \&
Cosmology}, {\bf 4}, 231.

	Noether, E. (1918). {\em G\"otting. Nachr.}, 235.

	Sakurai, J.J. (1960). {\em Ann. Phys. (New York)},
{\bf 11}, 1.

	Schutz, B.F. (1970). {\em Phys. Rev.},
{\bf D 2}, 2762.

	Taub, A.H. (1954). {\em Phys. Rev.}, {\bf 94},
1468.

	Tolman, R.C. (1934). {\em Proc. Nat. Acad. Sci. U. S.},
{\bf 20}, 169.

	Trautman, A. (1956). {\em Bull. Acad. Polon.
Sci., S\'er. III}, {\bf 4}, 665 \& 671.

	Weinberg, S. (1996). {\it The Quantum Theory of Fields.}
Vol. I: Foundations, Cambridge University Press, Cambridge.
\end{document}